\documentclass[12pt]{article}
\usepackage[pdftex]{graphicx} 
\usepackage{epsfig}
\usepackage{comment}
\usepackage{latexsym}
\usepackage{hyperref}
\usepackage{amsmath}
\usepackage{color}

\newcommand{\mysquare}[0]{\raise-.2ex\hbox{{\Large$\Box$}}}
\def\lsim{\mathrel{\rlap {\raise.5ex\hbox{$ < $}}
{\lower.5ex\hbox{$\sim$}}}}
\def\gsim{\mathrel{\rlap {\raise.5ex\hbox{$ > $}}
{\lower.5ex\hbox{$\sim$}}}} \topmargin -1.5cm \textheight=22.5cm \textwidth=16.5cm
\catcode`\@=11
\newcount\hour
\newcount\minute
\newtoks\amorpm
\hour=\time\divide\hour by60 \minute=\time{\multiply\hour by60
\global\advance\minute by-\hour}
\edef\standardtime{{\ifnum\hour<12 \global\amorpm={am}%
        \else\global\amorpm={pm}\advance\hour by-12 \fi
        \ifnum\hour=0 \hour=12 \fi
        \number\hour:\ifnum\minute<10 0\fi\number\minute\the\amorpm}}
\edef\militarytime{\number\hour:\ifnum\minute<10 0\fi\number\minute}
\def\draftlabel#1{{\@bsphack\if@filesw {\let\thepage\relax
   \xdef\@gtempa{\write\@auxout{\string
      \newlabel{#1}{{\@currentlabel}{\thepage}}}}}\@gtempa
   \if@nobreak \ifvmode\nobreak\fi\fi\fi\@esphack}
        \gdef\@eqnlabel{#1}}
\def\@eqnlabel{}
\def\@vacuum{}
\def\draftmarginnote#1{\marginpar{\raggedright\scriptsize\tt#1}}
\def\draft{\oddsidemargin -.2truein
        \def\@oddfoot{\sl preliminary draft \hfil
        \rm\thepage\hfil\sl\today\quad\militarytime}
        \let\@evenfoot\@oddfoot \overfullrule 3pt
        \let\label=\draftlabel
        \let\marginnote=\draftmarginnote
   \def\@eqnnum{(\theequation)\rlap{\k

 ern\marginparsep\tt\@eqnlabel}%
\global\let\@eqnlabel\@vacuum}  }

\newcommand{\be}[0]{\begin{equation}}
\newcommand{\ee}[0]{\end{equation}}
\newcommand{\ba}[0]{\begin{eqnarray}}
\newcommand{\ea}[0]{\end{eqnarray}}

\def\bs{\begin{subequations}}
\def\es{\end{subequations}}

\def\thebibliography#1{%
\vskip 0.5cm \centerline{\bf \Large References}
\list{%
[\arabic{enumi}]}{\settowidth\labelwidth{[#1]} \leftmargin\labelwidth
\advance\leftmargin\labelsep
\usecounter{enumi}}
\def\newblock{\hskip .11em plus .33em minus .07em}
\sloppy\clubpenalty4000\widowpenalty4000 \sfcode`\.=1000\relax}

\renewcommand{\theequation}{\arabic{section}.\arabic{equation}}

\renewcommand{\section}{\setcounter{equation}{0}\@startsection
{section}{1}{0mm}{-\baselineskip}{0.5\baselineskip} {\normalfont\Large\bfseries}}

\renewcommand{\subsection}{\@startsection
{subsection}{2}{0mm}{-\baselineskip}{0.5\baselineskip} {\normalfont\large\bfseries}}

\renewcommand{\subsubsection}{\@startsection
{subsubsection}{3}{0mm}{-\baselineskip}{0.5\baselineskip}
{\normalfont\normalsize\slshape}}

\usepackage{amssymb}
\usepackage{amsthm}
\usepackage{amsmath}
\usepackage{amssymb,amsfonts}
\usepackage{graphicx}
\usepackage{cite}

\renewcommand{\and}{\mbox{and}}

\begin{document}
\begin{titlepage}
\begin{flushright}
LPTENS--13/12,
May 2013
\vspace{-.0cm}
\end{flushright}
\begin{centering}
{\bf \Large Aspects of String Cosmology  }

\vspace{5mm}

 {\bf Costas Kounnas$^{1}$ and Nicolaos Toumbas$^{2\dag}$}

\vspace{1mm}

$^1$ Laboratoire de Physique Th\'eorique,
Ecole Normale Sup\'erieure,$^{\dag\dag}$ \\
24 rue Lhomond, F--75231 Paris cedex 05, France\\
{\em  Costas.Kounnas@lpt.ens.fr}

$^2$  Department of Physics, University of Cyprus,\\
Nicosia 1678, Cyprus.\\
{\em nick@ucy.ac.cy}

\end{centering}
\vspace{0.1cm}
$~$\\
\centerline{\bf\Large Abstract}\\
\vspace{-0.2cm}

\noindent 

$~$\\
We review recent progress in string cosmology, where string dualities are applied so as to
obtain complete cosmological evolutions, free of any essential singularities. Two classes of models are analyzed.
The first class consists of string gas cosmologies associated to certain thermal configurations of type II ${\cal N}=(4,0)$ models. 
Finite temperature is introduced along with non-trivial ``gravito-magnetic'' fluxes that lift the Hagedorn instabilities 
of the canonical ensemble and restore thermal T-duality symmetry. 
At a critical maximal temperature additional thermal states become massless sourcing stringy S-branes,
which facilitate a bounce between the two dual, asymptotically cold phases. 
Unlike previous incarnations of pre-Big Bang cosmologies,
the models remain perturbative throughout the cosmological evolution. 
The second class consists of exact solutions to classical 
string theory that admit a Euclidean description in terms of compact parafermionic worldsheet systems. 
The Euclidean target space corresponds to a non-singular, compact T-fold, 
which can be used to construct a normalizable Hartle-Hawking wavefunction for the cosmology.\\

$~$\\
\centerline{\it Lectures given  at the Corfu Summer Institute 2012}
\centerline{\it``School and Workshops on Elementary Particle Physics and Gravity"}
\centerline{\it September 8-27, 2012, Corfu, Greece.}


\vspace{3pt} \vfill \hrule \vskip.1mm{\small \small \small
  \noindent  $^{\dag}$ Speaker \\
  \noindent
   $^{\dag\dag}$\ Unit{\'e} mixte  du CNRS et de l'Ecole Normale Sup{\'e}rieure UMR 8549 \\ associ\'ee \`a
l'Universit\'e Pierre et Marie Curie (Paris 6), UMR 8549.}\\

\end{titlepage}
\newpage
\setcounter{footnote}{0}
\renewcommand{\thefootnote}{\arabic{footnote}}
 \setlength{\baselineskip}{.7cm} \setlength{\parskip}{.2cm}

\setcounter{section}{0}


\section{Introduction}

The hot Big-Bang model gives a robust description of the evolution of the Universe, from the onset of Nucleosynthesis until present \cite{KolbTurner}. 
Key cosmological 
puzzles concerning the observed large-scale smoothness, the flatness and horizon problems have inspired inflationary cosmology, 
where a phase of rapid accelerated expansion occurs in the very early cosmological history, eventually settling into radiation-dominated evolution \cite{Inflation}.      
The model can be adapted so as to take into account the current accelerating expansion of the Universe, 
as well as gravitational effects observed in large scale structures, by introducing dark energy and dark matter \cite{DE}. 
Thus the resulting cosmological scenario involves  a very rich phenomenological model, 
called the $\rm \Lambda CDM$ model,
based on classical general relativity and quantum field theory, with high temperature eras, symmetry breaking phase transitions, and 
proportionally large amounts of dark matter and dark energy dominating the very late time evolution.

This standard cosmological model presents some of the greatest challenges to fundamental physics. Two of these have proved to be particularly
acute over the years. Firstly, if we extrapolate the cosmological evolution back in time, using the equations of general relativity and 
quantum field theory, we are driven to an initial singularity, where the Universe collapses to zero volume and the description breaks down \cite{HE}.    
The second concerns the nature of dark energy. The simplest explanation for it is a positive, however unnaturally small, cosmological constant,
$\Lambda \sim 10^{-120} M_p^4 $, many orders of magnitude smaller than the Planck and elementary particle physics scales. 
To date no symmetry principle
or mechanism is known to explain its value. (For a recent review concerning the cosmological constant problem, see \cite{Polchinski}.) 
Moreover, if dark energy persists arbitrarily long, it would imply that the Universe approaches de Sitter space in the far future, with
a cosmic event horizon, and so portions of space will remain unobservable, forever. The observable part of the Universe is in a highly mixed
state. Therefore, within the context of general relativity and the Standard Model, we lack a coherent framework to analyze the cosmology of our Universe, from beginning to end.

If string theory is a complete theory of quantum gravity, it should eventually provide a consistent cosmological framework. The hope is that by incorporating fundamental duality symmetries and stringy degrees of freedom in time-dependent settings, we will be able to obtain complete cosmological histories, free of any essential singularities, and new tools for model building. 

Indeed, string dualities have given us profound insights into the nature of {\it Space} over the years. 
New phenomena arise at short distances 
of order the string scale, $l_s=\sqrt{\alpha'}$, or the Planck length, $l_p$, 
which do not admit a conventional field theory description, with Riemannian concepts breaking down.
Unlike a field theoretic incarnation, string theory is a UV finite theory of quantum gravity. T-duality, or small/large volume duality, 
implies that shrinking radii past
the string scale does not produce a lower dimensional theory. The stringy spacetime uncertainty principles, 
$\Delta x \Delta t \sim l_s^2$, 
$\Delta x \Delta t \sim l_p^2$, point to a minimal length and an intrinsic non-commutative geometry, 
leading to the UV/IR connection \cite{Y}. There are examples of singularity resolution, such as orbifold \cite{DHVW} and conifold singularities 
\cite{Stromingerconi}, and
topology change \cite{Witten1,CosmoTopologyChange}, where the appearance of extra massless states, localized at the singularity, make it fuzzy and smooth. 
Non-conventional thermodynamics, with Hagedorn and black hole phases, signal a maximal temperature and non-trivial phase transitions 
\cite{H,Kogan,Sathiapalan,AW,BG,AK,ADK,BR}. 
Finally, there are robust examples of non-perturbative strong/weak coupling dualities -- see e.g. \cite{Dualities} and references
therein -- and holographic gauge theory/gravity dualities \cite{Holo,Maldacena},
illustrating how string theory can provide concrete answers to many of the puzzles one has to face in trying to quantize Einstein's theory
of general relativity.

Some of the important lessons, relevant to the discussion below, are the following. Locality, geometry and even topology are approximate concepts, acquiring a more
precise definition at low enough energies: $E\le 1/l_s$ \cite{CosmoTopologyChange,Seiberg}. From holography, we learn that gravity and {\it Space} can emerge in special,
quantum mechanical systems. E.g. a large $N$ maximally supersymmetric gauge theory in $4$-dimensions gives rise to a $10$-dimensional
gravitational theory, a string theory on $AdS_5 \times S_5$ \cite{Maldacena}. Finally, apparently singular regimes and/or geometries can be mapped via
string dualities into non-singular ones, with well defined effective descriptions \cite{CosmoTopologyChange,Seiberg}. 
Much of this insight has been obtained from studies of static, equilibrium
configurations of superstrings, but given the principle of relativity, it seems inevitable that similar results hold for the time-dependent cases.

Thus progress in String Cosmology can be achieved if we manage to extend the web of string dualities to time-dependent, cosmological settings. See 
e.g. \cite{BV,TV,V,GV,BFM,KOSST}\cite{CosmoTopologyChange} for work towards this end.
This endeavor is both technically and conceptually challenging. Indeed, in phases with (spontaneous) supersymmetry breaking and geometric variation the 
string equations can become very difficult to solve, and in many cases even hard to formulate. With the moduli acquiring time-dependence, 
these may wander through cross-over regions of moduli space, where we have no
control over the quantum corrections, and the effective field theory approach breaks down. At a more fundamental level, it is hard to identify and compute
the correct, precise observables, and we lack a second quantized version of the theory to probe it directly off-shell.

It is also challenging to extend the web of holographic dualities to cosmological backgrounds. The construction of the holographic theory is very sensitive
to the global structure of spacetime, with the dual variables living at the boundary of spacetime. For asymptotically de Sitter cosmologies (like our own), the
natural boundaries lie to the infinite future, suggesting a form of spacelike holography: E.g. the $dS_4$/Euclidean $CFT_3$ correspondence \cite{Strominger}.
The dual holographic theory is conjectured to be a $3$-dimensional CFT living on the future (spacelike) boundary of de Sitter space. Symmetry considerations fix
the central charge of the CFT to be inversely proportional to the square of the asymptotic value of the Hubble parameter -- the Hubble parameter $H(t)$ decreases
monotonically with time and asymptotes to a
constant for asymptotically de Sitter cosmologies. Thus, $c \sim 1/ [H(t\to\infty)]^2G$. 

Such a boundary CFT, if it exists, may be non-unitary, and in fact it was
argued that the finiteness of de Sitter entropy implies the existence of Poincare recurrences at very late time scales, which in turn prevent the
realization of local observables in the infinite time limit \cite{DyLS}. There is no explicit, microscopic construction in string theory. If realized however, it would
be a holographic example where {\it Time} and the cosmological history emerge, perhaps as RG flow in the CFT. The relation between the central charge and the
Hubble parameter suggests that the field theory RG flow gets mapped to the time-reversed cosmological evolution. Therefore, reconstructing the very early Universe would
amount to the difficult task of decoding the hologram in the deep infrared of the boundary CFT. Another conceptual difficulty is that no single observer can measure
the boundary CFT correlators.  

In this lecture we will revisit the possibility of realizing eternal string cosmologies where an initially contacting phase bounces/emerges into  
an expanding thermal phase. 
We will argue that string theory contains the ingredients which could resolve strong curvature regimes, via duality transformations 
that lead to a well-defined, effective description \cite{BV,TV,V,GV,BFM,KOSST}\cite{CosmoTopologyChange}. 
Two classes of string cosmological solutions will be discussed to illustrate this pattern.
The first class consists of string gas cosmologies associated to certain special, thermal configurations of type II ${\cal N}=(4,0)$ models \cite{FKPT,KPT1,KPT2}. 
Finite temperature is introduced along with non-trivial ``gravito-magnetic'' fluxes that lift the Hagedorn instabilities of the canonical ensemble 
and restore thermal T-duality symmetry \cite{AKPT,massivesusy,FKT} \footnote{Additional work on thermal duality includes \cite{Chaud}.}. 
The cosmological evolutions describe bouncing Universes, with the bounce occurring at a stringy extended symmetry point.     
The second class consists of exact solutions to classical string theory that admit a Euclidean description in terms of compact parafermionic worldsheet systems \cite{KL,KTT}. 
The Euclidean target space corresponds to a non-singular, compact T-fold, which can be used to construct a normalizable Hartle-Hawking wavefunction for the cosmology \cite{KTT}.

\section{Bouncing string cosmologies}

Before focusing on stringy examples, we review the situation in classical general relativity. The singularity theorems of Penrose and Hawking show that a smooth reversal
from contraction to expansion is impossible unless an energy condition of the form 
\begin{equation}
T_{\mu\nu}v^{\mu}v^{\nu}\ge 0
\end{equation} 
is violated \cite{HE}. For the null energy condition (NEC), $v^{\mu}$ stands for any null future pointing vector. 
Let us see how these theorems apply for homogeneous and isotropic Friedmann-Robertson-Walker (FRW) cosmologies:
\begin{equation}
ds^2=-dt^2+[a(t)]^2d\Omega_k^2.
\end{equation}
As usual $a$ stands for the scale factor and the Hubble parameter is given by $H=\dot a /a$. 
We also assume that the cosmological evolution is supported by various sources with total energy density $\rho$ and pressure $P$,
which comprise together a perfect fluid.

The relevant equations are the Friedmann-Hubble equation
\begin{equation}
H^2 = {8\pi \over 3} G\rho - {k \over a^2}
\end{equation} 
and the 1st law of thermodynamics for adiabatic evolution:
\begin{equation}
\dot \rho +3H(\rho+P)=0.
\end{equation}
These two equations imply that
\begin{equation}
\dot H = -4\pi G(\rho +P)+{k \over a^2}.
\end{equation}
Therefore, if the NEC is satisfied,
\begin{equation}
\rho+P\ge 0,
\end{equation}
for flat ($k=0$) and open ($k=-1$) Universes, the Hubble parameter decreases monotonically with time, $\dot H\le 0$, and reversal from contraction ($H <0$) to
expansion ($H>0$) is not possible. The two phases are separated by a singularity, or the expanding phase is past geodesically incomplete. All known (macroscopic)
sources of energy and matter in our Universe satisfy the NEC.

For closed Universes ($k=1$), we get
\begin{equation}
\ddot a=-{4\pi \over 3}Ga(\rho+3P),
\end{equation}  
and so for the bounce to occur at a singularity the strong energy condition (SEC),
\begin{equation}
\rho \ge -3P,    
\end{equation}    
must be satisfied. There are field theoretic sources for which the NEC holds but the SEC is violated, e.g. positive vacuum energy or a positive
cosmological constant, and global de Sitter space is an example of a closed FRW cosmology, where (at the classical level) a contracting phase smoothly   
reverses to an expanding inflationary phase. The problem with such an initially exponentially contracting phase is that it requires the Universe to be sufficiently empty for an infinite amount of time. During exponential contraction perturbations grow large, and so the Universe is likely to thermalize before expansion sets in, and within the field theoretic context, collapse to a singularity. See \cite{Bousso} for further discussions. 

Another possibility is to consider a Universe which is eternally inflating, or expanding sufficiently fast forever (requiring that $H_{av}>0$ throughout
the cosmological history). Even without requiring an energy condition, the authors of \cite{BGV} show that such a Universe cannot be past geodesically complete and must have a beginning, presumably an initial singularity. Scalar field driven inflation cannot be the ultimate theory of the very early Universe. 

There are various notable attempts to overcome the reversal problem within the string theoretic set-up. Let us summarize the main ideas of some of these.
\begin{itemize}
\item There are various incarnations of the pre Big Bang scenario \cite{GV}. Typically in the pre Big Bang phase the dilaton runs from weak to strong coupling. As we approach strong coupling, new terms in the effective action such as higher derivative interactions and potentials can become relevant, invalidating some of the assumptions of the singularity theorems and thus facilitating the bounce. The difficulty of these models is in maintaining analytical control over the strongly coupled dynamics at the bounce.
\item Various (weakly coupled) null/spacelike orbifold models of the singularities, where the orbifold is obtained by modding out with a boost \cite{CC,LMS}. 
Adding a particle in such a spacetime, amounts to also adding in the covering space an infinite number of boosted images, causing strong backreaction, and
possibly gravitational collapse \cite{HP1}. See also \cite{CC2} for a counterexample.
\item String gas cosmology \cite{BV,TV}. Here the idea is that the Universe starts as a compact space, e.g. a $9$-dimensional torus with all radii close to the string scale and temperature close to (but below) the Hagedorn temperature: $T \sim T_H$. There is a competition between the thermally excited momentum modes, which keep the spatial torus from shrinking, and thermally excited winding modes which prevent the Universe from expanding. The system is presumed to be in a quasi-static phase, until thermal fluctuations cause the winding states to annihilate and some dimensions to expand. At the level of the two derivative dilaton/gravity effective action, there is a singularity a finite time in the past, where also the dilaton grows to strong coupling.
\item Modular cosmology \cite{HM,BFM}, cosmological billiards \cite{DN}, brane collisions and the Ekpyrotic scenario \cite{KOST} and matrix cosmology \cite{SV}.    
\end{itemize}  
Further work on string cosmology includes \cite{HotstringsCosmo}.

We shall focus on another class of bouncing thermal string cosmologies, exploiting stringy phase transitions and phenomena that can occur at temperatures close to the 
Hagedorn temperature. The non-singular cosmological solutions are based on a mechanism that resolves the Hagedorn instabilities of finite temperature strings,
realizable in a large class of initially ${\cal N}=(4,0)$ type II superstring models \cite{FKPT,KPT1,KPT2}. 
We also review some aspects of strings at finite temperature and (partial) spontaneous 
supersymmetry breaking via geometrical fluxes.

\subsection{Thermal configurations of type II ${\cal N}=(4,0)$ models}

We consider weakly coupled type II $(4,0)$ models on initially flat backgrounds:
\begin{equation}
R^d \times T^{10-d},
\end{equation}
where the internal toroidal radii are taken close to the string scale.
There are $16$ real spacetime supersymmetries arising from the left-moving sector of the worldsheet. 
The right-moving susy is broken spontaneously by twisting some of the internal radii with $F_R$, the right-moving fermion number. 
Under the $Z_2$ symmetry $(-1)^{F_R}$ the
right-moving $R$ sector changes sign.

This pattern of asymmetric susy breaking leads to extended symmetry points, when the internal radii are 
at the fermionic point \cite{FKPT,KPT1,KPT2,AKPT,massivesusy,FKT}. 
At finite temperature, such points in moduli space are preferred, with the moduli participating in the breaking of the right-moving supersymmetries being
stabilized at the extended symmetry point values \cite{stabmod}. 
As a result the odd-$F_R$ sector is heavy, with masses being bounded from below by the string scale: $m^2 \ge 1/(2\alpha')$.

As illustrative examples, consider the two dimensional Hybrid vacua, on $R^2\times T^8$, where all the internal radii are taken at the fermionic point
$R=1/\sqrt{2}$ \cite{FKT,FKPT} -- we work in string units where $\alpha'=1$. At this point the eight compact supercoordinates can be replaced with 24 left-moving and 
24 right-moving worldsheet fermions. The 24 left-moving fermions are split into two groups of 8 and 16. The one-loop partition function is given by
\begin{equation}
Z_{\rm Hyb}={V_2\over (2\pi)^2}~\int_{\cal{F}} {d^2\tau\over
4({\rm Im}\tau)^{2}}~{1 \over \eta^8}~\Gamma_{E_8}(\tau)~\left(V_8 - S_8\right)~\left(\bar V_{24}-\bar S_{24}\right),
\end{equation}  
and exhibits holomorphic/anti-holomorphic factorization.

In the left-moving sector, the group of 8 fermions are described in terms of the $SO(8)$ characters,
$$
O_8 = \frac{\theta_3^4 + \theta_4^4 }{ 2\,\eta^4}\,,
\qquad 
V_8 = \frac{\theta_3^4 - \theta_4^4}{ 2\, \eta^4}\,,
$$
\begin{equation}
S_8  = \frac{\theta_2^4 - \theta_1^4}{ 2\,\eta^4}\,,
\qquad
C_8= \frac{\theta_2^4 + \theta_1^4 }{2\, \eta^4}\,,
\end{equation}
as in the conventional superstring models, and the other 16 fermions are described by the chiral $E_8$ lattice: $\Gamma_{E_8}$.   
In the right-moving sector, the fermions are described in terms of the $SO(24)$ characters:
\begin{equation}
\bar V_{24}-\bar S_{24}={1\over 2 \bar\eta^{12}}\left(\bar\theta_3^{12} - \bar\theta_4^{12}-\bar\theta_2^{12}\right)=24. 
\end{equation}
Despite the breaking of the right-moving supersymmetries, this sector exhibits {\it Massive Spectrum Degeneracy Symmetry} (MSDS) \cite{massivesusy}.
This degeneracy is broken at the right-moving massless sector, a fact that leads to the breaking of the right-moving supersymmetries.
The left-moving supersymmetry remains unbroken.
The massless sector of the physical, level-matched, spectrum consists of $24 \times 8$ bosons and $24\times 8$ fermions arising in the 
$V_8 \bar V_{24}$ and $S_8\bar V_{24}$ sectors respectively. Notice in particular that the right moving $R$ sector is massive. 

The model can be also exhibited as a freely acting, asymmetric orbifold compactification of the type II superstring to $2$ dimensions.
The relevant half-shifted $(8,8)$ lattice is given by
$$
\Gamma_{(8,8)}\left[^{\bar a}_{\bar b}\right]=\Gamma_{E_8}~\times~\bar\theta\left[^{\bar a}_{\bar b}\right]^8 ~ \to
$$
\begin{equation}
{\sqrt{\det G_{IJ}}\over (\sqrt{\tau_2})^8}\sum_{\tilde m^I,n^J}e^{-{\pi \over \tau_2}(G+B)_{IJ}(\tilde m+\tau n)^I(\tilde m+\bar \tau n)^J}~\times~{e^{i\pi(\tilde m^1\bar a+n^1\bar b+\tilde m^1n^1)}}.
\end{equation}
The modular covariant cocycle describes the coupling of the lattice to the right-moving fermion number $F_R$. In particular, only one internal cycle is twisted
by $F_R$. At the MSDS point, the metric and antisymmetric B-field tensors, $G_{IJ}, \, B_{IJ}$ take special values, leading to holomorphic/anti-holomorphic factorization and enhanced gauge symmetry with
the local gauge group given by \cite{FKT} 
\begin{equation}
U(1)_L^8 \times [SU(2)_R]_{k=2}^8.
\end{equation}

We have focused on the highly symmetric Hybrid vacua, where the presence of exact right-moving MSDS symmetry 
leads to exact computations, as we will see below, but a large class of $(4,0)$ models can be constructed in various dimensions \cite{AKPT,KPT1,KPT2}. 

Next we consider the models at finite temperature. To avoid strong Jeans instabilities and gravitational collapse into black holes, we compactify $R^{d-1}$ on a large torus with each cycle having radius $R\gg 1$, and take the string coupling to be sufficiently weak. Backreaction can be ignored, if
the size of the thermal system is much larger than its Schwarzschild radius: $R_S \sim GM=G\rho R^3$ in 4 dimensions, where the energy density is set by the temperature.
This allows for the following range for the sizes of the radii $R$:
\begin{equation}
1\ll R \ll {1 \over \sqrt{G\rho}}\sim{1 \over g_s},
\end{equation}
where the last equality follows for temperatures close to the string scale. Both inequalities can be satisfied at sufficiently weak coupling. At larger values of the
coupling constant, we cannot ignore backreaction and we must take into account the induced cosmological evolution. The string coupling must still be kept
small so as to be able to maintain conditions of quasi-static thermal equilibrium and trust the perturbative computations of various thermodynamical quantities.

In string theory new instabilities set in at temperatures close to the string scale, $T \sim 1/l_s$, the Hagedorn instabilities, which 
signal non-trivial phase transitions \cite{Kogan,Sathiapalan,AW,BG,AK,ADK,BR}. 
The origin of these instabilities is due to the exponential rise in the density of (single-particle) string states at
large mass \cite{H}:
\begin{equation}
n(m) \sim e^{\beta_H m}.
\end{equation}
Because of the exponential growth, the single string partition function 
\begin{equation}
Z\sim\int^{\infty} dm~ n(m)e^{-\beta m}=\int^{\infty} dm~ e^{-(\beta -\beta_H)m}
\end{equation}
diverges for temperatures above Hagedorn: $T > T_H=1/\beta_H$. 
The Hagedorn temperature is set by the coefficient of the exponent in the asymptotic formula
for the density of states and it is close to the string scale: 
\begin{equation}
\beta_H=2\pi \sqrt{2\alpha'}
\end{equation}
in Type II superstrings.
  
Therefore the critical behavior as $\beta \to \beta_H$ is governed by string states of large mass $m \gg 1/l_s$, 
or high level $N$ -- recall that $m^2 \sim N/\alpha'$. 
At weak coupling, the typical size of such a string state is large,
of order $l \sim N^{1/4}l_s$ \cite{MT,HP2}. We can think of it as a random walk of $N^{1/2}$ bits.
Now the entropy carried by an excited long string of mass $m$ is greater than the entropy of $n$ smaller strings,
each having mass $m/n$.
So close to the Hagedorn temperature, percolation phenomena take place 
with multiple strings coalescing into fluctuations of a single long, tangled string.

The critical point can be also described by an effective field theory of a 
massless complex scalar field, manifesting a UV/IR connection \cite{Kogan,Sathiapalan,AW,AK,ADK}. In quantum field theory, the thermal effective theory
is obtained by compactifying Euclidean time on a circle with period set by the inverse temperature, $2\pi R_0 = 1/T$, 
and imposing periodic
boundary conditions for bosonic fields and anti-periodic boundary conditions for fermions.
In string theory, the thermal system can be described in terms of a freely acting orbifold, obtained by twisting the Euclidean time circle
with the spacetime fermion number $F$.
For type II superstrings this amounts to coupling the Euclidean time $\Gamma_{1,1}(R_0)$ lattice with the following co-cycle \cite{RK}:
\begin{equation}
e^{i\pi(\tilde m^0(a + \bar a)+ n^0(b+\bar b))}.
\end{equation}

In this picture, the instabilities appear at a critical compactification radius set by the Hagedorn temperature.
Certain string winding modes, with $(n_0\ne 0)$, become massless
precisely when $R_0 = R_H=1/(2\pi T_H)$. 
They become tachyonic at smaller radii, when $R_0 < R_H$. More precisely, two winding modes pair up to form a complex scalar field, whose
thermal mass is given by
\begin{equation}
m^2(R_0)=R_0^2 - R_H^2.
\end{equation}
As a result, near the critical point the behavior of the partition function is captured by the thermal scalar path integral:
$$
{\cal Z}\sim\int [d\varphi] e^{-S[\varphi]}
$$
\begin{equation}
S[\varphi] \sim \int d^{d-1}x (\partial_i \varphi^*\partial^i\varphi+m^2(R_0)\varphi^*\varphi).
\end{equation}
We can illustrate an aspect of this correspondence, by recovering the asymptotic formula for the density of states at large mass \cite{BV,HP2,AEK}. 
With all spatial dimensions being compact and close to the critical point, the logarithm of the partition function is 
dominated by the lowest eigenvalue of the Klein-Gordon operator $-\nabla^2+m^2(R_0)$, given by the square of the mass: $\lambda_0 = m^2(R_0)$.
Therefore
\begin{equation}
Z_c =\ln{\cal Z} \sim -\ln \lambda_0 \sim -\ln (R-R_H)\sim \int^{\infty} dm e^{-\beta m}n(m).
\end{equation}
The logarithmic behavior as $R \to R_H$ can be reproduced for 
\begin{equation}
n(m)=e^{\beta_H m}/m. 
\end{equation}
When $d-1$ spatial dimensions are non-compact, a similar computation yields \cite{BV,HP2,AEK}  
\begin{equation}
n(m)\sim V_{d-1} e^{\beta_H m}/m^{(d+1)/2}.
\end{equation} 

So the Hagedorn divergence for $R_0<R_H$ can be interpreted as an IR instability 
of the underlying Euclidean 
thermal background.
Tachyon condensation gives a genus-zero contribution to the free energy,
$F \sim 1/g_s^2$, 
leading to large backreaction, which,
presumably, brings the thermal ensemble to a speedy end \cite{AW}. 
In type II $(4,0)$ models, perturbatively stable configurations can be produced, 
if in addition to temperature, we
turn on vacuum potentials associated to the graviphoton $G_{I0}$ and $B_{I0}$ fields, where the index
$I$ is along an internal direction twisted by $F_R$ \cite{AKPT,FKT}. See also \cite{DiS}.
In particular, we turn on the $U(1)_L$ combination, $G_{0I}+2B_{0I}$, of these fields.
At finite temperature, such vacuum potentials cannot be gauged away, as they correspond to topological vacuum parameters.
These {\it gravito-magnetic} fluxes modify the thermal masses of all states charged under 
the graviphoton fields, and for large enough values, the  
tachyonic instabilities can be lifted.
Equivalently the contribution to the free energy of the massive oscillator states gets 
regulated (refined), 
reducing the effective density of thermally excited states, and restoring asymptotic supersymmetry \cite{KS}.

The Hagedorn free models can be described in terms of freely acting asymmetric orbifolds of the form $(-1)^{F_L}\delta_0$, where
$\delta_0$ is a $Z_2$-shift along the Euclidean time circle \cite{FKPT,KPT1,AKPT,FKT}. In the Hybrid example, 
the partition function is
given explicitly by
$$
{Z_{\rm Hyb} \over V_1} = \int_{\cal F} \frac{d^2 \tau}{8\pi({\rm Im}\tau)^{3/2}}~(\bar V_{24}- \bar S_{24}) \, 
~{\Gamma_{E_8}(\tau)\over \eta^8} 
$$
\begin{equation}
\times \sum_{m, n}
\left(V_8 \, \Gamma_{m,2n}(R_0) + O_8 \, \Gamma_{m +\frac{1}{2},2n +1}(R_0) - S_8 \, 
\Gamma_{m +\frac{1}{2} , 2 n}(R_0) -C_8\, \Gamma_{m, 2n +1}(R_0) \right),
\end{equation}
and it is finite for all values of the thermal modulus $R_0$. In fact, the model
remains tachyon-free under all deformations of the dynamical moduli associated with the compact, internal
eight-manifold \cite{FKT}.

All such models exhibit a number of universal properties, irrespectively of spacetime dimension \cite{KPT1}.
The gravito-magnetic fluxes lead to a restoration of the stringy T-duality symmetry along the thermal circle:  
\begin{equation}
R_0 \to R_c^2/R_0,~~~~ S_8 \leftrightarrow C_8,
\end{equation}
where in all models, the self-dual point occurs at the fermionic point $R_c=1/\sqrt{2}$.
The partition function is finite and duality invariant, but it is not a smooth function of $R_0$.  
At the self-dual point $R_c$ additional thermal states become massless enhancing the gauge symmetry associated to the Euclidean time circle,
\begin{equation} 
U(1)_L \times U(1)_R \to [SU(2)_L]_{k=2} \times U(1)_R,
\end{equation}
and inducing a conical structure in $Z$ (as a function of $R_0$), signaling a stringy phase transition. 
   
For example in the Hybrid model, $2 \times 24$ states in the $O_8 \bar V_{24}$ sector become massless, precisely at the self-dual point:
\begin{equation}
\label{mass}
m^2=\left({1\over 2R_0} -R_0\right)^2. 
\end{equation} 
Away from the critical point the mass square is strictly positive, and the states do not become tachyonic at smaller values of the
radius $R_0$ \footnote{This is to be contrasted with heterotic strings at finite temperature, where the two dual phases, at small and large values of 
the thermal modulus $R_0$, are separated
by an intermediate tachyonic region.}.
The corresponding left-moving and right-moving momentum charges and vertex operators are given by
\begin{equation}
\label{vertex}
p_L=\pm 1,~~~p_R=0,~~~O_\pm=\psi_L^0\, e^{\pm i X_L^0}\, {\cal O}_R.
\end{equation} 
So the additional massless states carry both non-trivial momentum and winding charges.

In the Hybrid model the thermal partition function can be computed exactly thanks to right-moving MSDS symmetry \cite{FKPT,FKT}: 
\begin{equation}
{Z_{\rm Hyb} \over V_1}=24\times \left(R_0 + {1 \over 2R_0}\right)-24 \times \left| R_0 - {1 \over 2R_0}\right|.
\end{equation}
There is
complete suppression of the massive oscillator contributions away from the critical point. However, stringy behavior survives  
at the critical point giving rise to the conical structure. From the thermal effective field theory point of view, such non-analytic behavior is induced 
after integrating out the additional massless states. With one spatial dimension non-compact, each complex boson
becoming massless contributes a factor given by the absolute value of the mass: $-|m|$. Since in the Hybrid model there are $24$ such states, with
masses given by equation (\ref{mass}), the non-analytic term in the partition function is accounted for. Thermal configurations of 
non-critical heterotic strings in two dimensions enjoy very similar properties \cite{DLS}.  

With $d-1$ spatial dimensions non-compact, the partition function acquires a higher order conical structure:
\begin{equation}
\sim \left| R_0 - {1 \over 2R_0}\right|^{d-1},
\end{equation}
implying a milder transition as a function of $R_0$.
Recall however that to avoid non-perturbative Jeans instabilities, we must keep all but at least one of the large spatial dimensions compact, 
and so the infinite volume and $R_0\to R_c$ limits may not commute. Therefore, we can take a number of spatial dimensions to be arbitrarily large (but compact), and
still the conical structure be linear at the critical point.  

Thermal duality implies the existence of two dual asymptotic regimes
dominated by the light thermal momenta, $R_0 \gg R_c$, and  
the light thermal windings, $R_0 \ll R_c$, respectively.
In the regime of light thermal momenta the partition function is given by
\begin{equation}
{Z \over V_{d-1}} ={n^* \Sigma_d\over (2\pi R_c)^{d-1}}\left({R_c \over R_0}\right)^{d-1}
+\;{\cal O}\left ( e ^{-R_0/R_c}  \right),
\end{equation}
giving rise to the characteristic behavior of massless thermal radiation in $d$ dimensions. The temperature is given by 
the inverse period of the Euclidean time circle, $T=1/2\pi R_0$; 
$n^*$ is the number of effectively massless degrees of freedom and $\Sigma_d$ stands for the Stefan-Boltzmann constant of massless
thermal radiation in $d$ dimensions.

By duality, we get that in the regime of light thermal windings, $R_0 \ll R_c$, the partition function is given by 
\begin{equation}
{Z \over V_{d-1}} ={n^* \Sigma_d\over (2\pi R_c)^{d-1}}\left({R_0 \over R_c}\right)^{d-1}
+\;{\cal O}\left ( e ^{-R_c/R_0}  \right).
\end{equation}
Notice in particular that $Z \to 0$ as $R_0 \to 0$. 
Now in standard thermodynamics the thermal partition function decreases monotonically as the temperature decreases.
So the correct definition of temperature cannot be $T=1/2\pi R_0$ in this regime. That is, the temperature in this regime is not set
by the inverse period of the Euclidean time circle.
The light winding excitations are non-local in $X^0$, but are local in the T-dual of $X^0$.  
In fact by T-duality, we can interpret them as ordinary 
thermal excitations associated with the large T-dual circle, whose radius is given by 
$\tilde R_0 = R_c^2/R_0$.
So the temperature in this regime is given by
$T= 1/2\pi\tilde R_0=R_0/(2\pi R_c^2)$, and the system at small radius $R_0$ is again effectively cold.

Thus the thermal system has two dual, asymptotically cold phases as the value of the thermal modulus $R_0$ varies.
Each thermal phase arises via spontaneous symmetry breaking, as we deform away from the 
intermediate extended symmetry point. 
The two phases are distinguished by the light thermally excited spinors. 
At large radii, $R_0>R_c$, these transform under the $S_8$-Spinor of the $SO(8)$ symmetry group, while at small radii,   
$R_0<R_c$, the light thermally excited spinors transform in terms of the conjugate $C_8$-Spinor. 
The extended symmetry point is purely stringy; it has no precise thermal interpretation but instead 
is T-duality invariant.

Operators associated with the extra massless states induce transitions
between the purely momentum and winding modes. 
The operators 
$O_+$ and $O_-$, given in equation (\ref{vertex}), raise and lower $p_L$ by one unit but leave $p_R$ unchanged. In particular since they
transform in the vector representation of the symmetry group $SO(8)$, 
they induce transitions between the purely momentum $S_8$ and purely winding $C_8$ spinors,
which become light in the two asymptotically cold regimes respectively:
\begin{equation}
<C_8|O_-|~S_8> \ne 0.
\end{equation}

As a result the stringy phase transition at $R_c$ can be resolved in the presence of genus-zero condensates 
of the additional massless thermal states, which can mediate transitions between purely winding and momentum states,
{\it ``gluing together'' the asymptotic regimes} \cite{FKPT,KPT1,KPT2}. These condensates can be described in terms of non-trivial textures, which define embeddings of the spatial
manifold into the field configuration space 
via non-zero
spatial gradients of the fields associated to the extra massless states:
$\nabla_\bot \varphi^I \ne 0$ \cite{KPT1}. As we will argue, in the Lorenztian description, where the temperature becomes time dependent, the 
condensates give rise to a 
spacelike brane (S-brane) configuration with negative pressure contributions localized at a time slice at which  
the temperature reaches its critical value.

Therefore thermal duality implies the existence of a maximal critical temperature $T\le T_c$.
The stringy system conceals its short distance behavior. Defining the thermal modulus $\sigma$ via
$R_0/R_c = e^\sigma$, T-duality acts by reversing its sign: $\sigma \to -\sigma$.
In terms of $\sigma$ and the critical temperature $T_c$, the physical temperature can be written in a duality invariant way as follows:
\begin{equation}
T=T_c~ e^{-|\sigma|}, ~~~T_c={1 \over 2\pi R_c}= {1 \over \sqrt{2}\pi}.
\end{equation}
The expression is valid in both asymptotically cold regimes.
Consequently the energy density and pressure are bounded
$\rho \le \rho_c$, $P \le P_c$,   
never exceeding certain maximal values. This is a
crucial difference from thermal field theory models.

In the Hybrid model, the thermal partition function can be written as
\begin{equation}
{Z \over V_1}=(24\sqrt{2})~e^{-|\sigma|}=\Lambda~{T },\,\,\,\,~~ 
\Lambda={24\sqrt{2}\over T_c}.
\end{equation}
and so the pressure, energy density and maximal energy density are given by      
\begin{equation}
P=\rho= \Lambda~T^2,~~~~\rho_c=\Lambda~T_c^2={24 \over \pi}.
\end{equation}
So in each thermal phase the equation of state is effectively that of thermal massless radiation 
in two dimensions thanks to right moving MSDS symmetry. 
In the higher dimensional cases, the exact right-moving MSDS structure gets replaced by  
{\it right moving asymptotic supersymmetry},
ensuring that well up to the critical point, the partition function is dominated by the contributions of the
thermally excited massless states: $Z \sim T^{d-1}$ \cite{KPT1}.
In order to maintain semi-quantitative control in the remaining part of this work, we will ignore stringy corrections to the one-loop
string partition function in the higher $d$ cases close to the critical point, and approximate the thermal system with that of massless thermal radiation
up to the critical temperature. We will incorporate however the crucial localized pressure contributions induced by the additional massless thermal states
at the critical point. 

\subsection{Cosmological evolutions} 
The backreaction on the
initially flat metric and dilaton background will induce a
cosmological evolution.
We consider first the case where the underlying thermal modulus $\sigma$ is a monotonic function of time  
scanning all three regimes of the string thermal system \cite{FKPT,KPT1}. Correspondingly the temperature grows from small values,
reaching its maximal value $T_c$, and then drops again to zero.  
Each regime admits a local effective
theory description, associated with a {\it distinct $\alpha'$-expansion:}
\begin{itemize}
\item $R_c/R_0 \gg 1$, $(\sigma \ll 0)$; the regime of light thermal windings: $\{{\cal W}(\sigma <0)\}$.
\item $\left| R_0/R_c - R_c/R_0\right|\ll 1,$ $(\sigma \sim 0)$; the intermediate $SU(2)$ extended symmetry point, where additional thermal
states become massless: $\{{\cal B}(\sigma =0)\}$.
\item $R_0/R_c \gg 1,$ $(\sigma \gg 0)$; the regime of light thermal momenta: $\{{\cal M}(\sigma >0)\}$.
\end{itemize}

So a stringy transition occurs at $T_c$ $(\sigma=0)$, connecting two asymptotically cold phases. In each of these phases the 
source comprises a thermal gas of strings coupled to the dilaton/gravity system.
Near $T_c$ condensates associated with the extra massless thermal scalars can form and decay giving rise to an 
S-brane configuration \footnote{It would be interesting to associate this configuration with formation and decay of long, tangled strings spread in space.} --
see \cite{SG} for discussions concerning stringy S-branes in various contexts.
Precisely at the critical point, the equations of motion of the thermal effective theory allow for non-trivial 
(genus-0) backgrounds, in which the gradients of the fields associated to the extra massless thermal states satisfy \cite{KPT1}
\begin{equation}
\label{gradients}
G_{IJ}\nabla_{\hat \mu}\varphi^I\nabla_{\hat \nu} \varphi^J = {\kappa \over (d-1)}g_{\hat \mu \hat \nu},
\end{equation}
where 
$G_{IJ}$ is the metric on the field configuration space; $g_{\hat \mu \hat \nu}$ is the metric on the spatial manifold and 
$\kappa$ is a positive
constant that sets the strength of the condensates. The resulting stress tensor is compatible with the symmetries of the
spatial metric, which we require to be homogeneous and isotropic.
Effectively, the condensates amount to {\it negative pressure contributions, proportional to $-\kappa$, which are 
well localized around the transition surface $T=T_c$ in the Lorentzian}.

At finite string coupling, we expect the thickness or duration of the S-brane in time to be very short, set by the string scale. 
So in the Lorentzian effective action 
we treat the S-brane as a $\delta$-function source (thin-brane limit). (Later on we will explore the possibility of spreading the intermediate regime, 
with the temperature being constant at its critical value, for a long period in time.) Integrating out the extra massless thermal scalars via equation (\ref{gradients}),
we obtain the brane contribution to the effective action \cite{FKPT,KPT1}: 
$$
S_{\rm brane}=-\kappa \int d\sigma d^{d-1}x \sqrt{{g_\bot}}e^{-2\phi}\delta(\sigma) \to
$$ 
\begin{equation}
-\kappa \int d\tau d^{d-1}x \sqrt{{g_\bot}}e^{-2\phi}\delta(\tau-\tau_c),  
\end{equation}  
where $\tau_c$ is the time at which the temperature reaches its critical value, $\sigma(\tau_c)=0$, and
$g_\bot$ stands for the determinant of the induced metric on the constant time slice $\tau=\tau_c$. 
The tension of the brane is set by $\kappa$. The energy density of such a spacelike brane vanishes 
(consistently with reflecting boundary conditions 
on the first time derivative of the dilaton field \cite{FKPT}), but it gives negative pressure contributions in
the spatial directions. Thus it provides the violation of the null energy condition (NEC) which can lead to a transition 
from a contracting phase to an expanding
phase (when the cosmology is viewed in the Einstein frame). Henceforth we will refer to the intermediate regime as the ``Brane Regime''.  

The first class of non-singular string cosmologies we discuss consists of transitions from a ``Winding regime'' to a ``Momentum regime'' 
via such a thin S-brane:    
\begin{equation}
{\cal C}(\tau) \equiv  \{{\cal W}(\tau<\tau_c)\}  \oplus \{{\cal B}(\tau=\tau_c)\} \oplus \{{\cal M} (\tau>\tau_c)\}.
\end{equation}
Without loss of generality we may choose $\tau_c=0$. Since we take the string coupling to be weak and the
size of the spatial manifold to be much larger than the string scale, our discussions above lead to the following effective action for the
cosmological dynamics, which is valid in both the asymptotic regimes and also close to the critical point: 
\begin{equation}
\label{action}
S = \int{d^d x~ e^{-2\phi}~\sqrt{-g}
\left(\frac{1}{2}~R+2(\nabla\phi)^2\right)}
+
 \int{d^d x~ \sqrt{-g}~ P} 
-\kappa \int d^{d}x \sqrt{g_{\bot}}e^{-2\phi}\delta(\tau).
\end{equation}
The first term is the genus-0 dilaton-gravity action written in string frame, while the second 
is the contribution of the thermal effective potential $-P$.
In terms of the genus-1 thermal partition function, the energy density and pressure are given by
\begin{equation}
P=T~{Z\over V_{d-1}}, ~~~~~~\rho=-P +T{\partial P \over \partial T}\sim (d-1)P.
\end{equation} 
The equation of state $\rho =P$ is exact in the 2d Hybrid models. 

We concentrate on spatially flat homogeneous and isotropic solutions 
\begin{equation}
ds^2=-N(\tau)^2{d\tau}^2+a(\tau)^2dx_idx^i, 
\end{equation}
where $a$ is the string frame scale factor and $N$ is the lapse function. Homogeneous and isotropic solutions with
negative spatial curvature can be also constructed.
Since in both thermal phases the Universe is asymptotically cold, this must be a bouncing 
cosmology. Irrespectively of the running dilaton, 
the thermal entropy (per co-moving cell of unit coordinate volume, physical volume $a^{d-1}$)
\begin{equation}
\label{entropy1}
S= {a^{d-1} \over T}(\rho + P)\sim (a T)^{d-1}
\end{equation}
is conserved, implying that in each of the two asymptotic thermal phases
the scale factor and temperature satisfy:
\begin{equation}
a T = \rm{constant}.
\end{equation}  
Therefore, during the initial ``Winding Regime'' where the temperature increases, the Universe is in a
contracting phase, which reverses to expansion once the Universe enters into the ``Momentum Regime'' with decreasing temperature.
Moreover, since the temperature and all thermodynamical quantities are bounded from above by critical values, which they attain at the
brane, the scale factor must be bounded from below by its value at the brane: 
$a \ge a_c$. 
This critical value of the scale factor is  
given in terms of the entropy and the maximal critical temperature: 
\begin{equation}
a_c = \left( S T_c \over \rho_c +P_c\right)^{1/d-1}\sim {S^{1/d-1} \over T_c}.
\end{equation}
In particular $a_c$ can be kept large in string units if the entropy $S$ is large.
The bounce is facilitated by the extra negative pressure provided by the
S-brane at the transition surface, inducing also a bounce on the dilaton.
The singular regime $a \to 0$ of classical general relativity is absent.

Notice that continuity of the scale factor across the brane, ensures the continuity of the thermal
entropy across the transition surface, as in second order phase transitions. The reversal of contraction to
expansion via the S-brane is what makes the transition between the two asymptotically cold phases possible,
avoiding a ``heat death'' and maintaining adiabaticity throughout and across the brane.

From the action (\ref{action}), we can obtain the equations of motion \cite{KPT1}. Varying with the lapse function $N$, we
get the following first order equation
\begin{equation}
\label{N}
{1\over2}(d-1)(d-2) H^2=2(d-1)H\dot \phi-2\dot\phi^2+e^{2\phi}N^2\rho,
\end{equation}
consistently with the vanishing of the brane contribution to the energy density.
For the scale factor $a$, we get
$$
(d-2)\left({\ddot a\over a} - H{\dot N\over N}\right)+{1\over2}(d-2)(d-3) H^2 
$$
\begin{equation}
= 2\ddot\phi+2(d-2)H\dot\phi-2\dot\phi^2-2{\dot N\over N}\dot\phi-e^{2\phi}N^2 P+ \kappa N\delta(\tau),
\end{equation}
including the localized negative pressure contributions from the S-brane. Finally the equation of motion of the dilaton
is
\begin{equation}
{\ddot\phi}+(d-1)H\dot\phi-\dot\phi^2-{\dot N\over N}\dot\phi {-{d-1\over 2}}\left({{\ddot a\over a}} - H{\dot N\over N}\right)
-{1\over 4}(d-1)(d-2) H^2
=-{1\over 2}\kappa N \delta(\tau). 
\end{equation}

Irrespectively of spacetime dimension, the structure of these equations is such that the dilaton experiences
an impulsive force at the brane,
$2\ddot \phi = -\kappa N_c \delta(\tau) + \dots$, while $\ddot a$ is smooth.
So the brane induces a discontinuity in the first time derivative of the dilaton field 
\begin{equation}
2(\dot \phi_+ -\dot \phi_-)=-N_c \kappa,
\end{equation}
while the first time derivative of the scale factor $\dot a$ remains continuous.
Criticality of the temperature at the brane -- this is a crucial string theory input -- leads to the vanishing of
the first time derivative of the scale factor: $\dot a =0$. Continuity of the dilaton field and the metric across the brane, 
and the first order equation (\ref{N}) then imply that
$\dot \phi_+=-\dot \phi_-$. In other words, the dilaton undergoes an {\it elastic} bounce across the brane:
\begin{equation}
\dot \phi_+=-\dot \phi_-=-N_c \kappa/4.
\end{equation}
Since the brane tension $\kappa$ is positive, 
the dilaton must be initially increasing. It crosses the brane and then decreases. As a result, the dilaton is bounded from above by its
value at the brane: $\phi \le \phi_c$. The critical value of the dilaton and the maximal energy density set 
the slope of the dilaton just before the transition, the brane tension $\kappa$, and hence the strength of the condensates
associated to the extra massless thermal scalars, via equation (\ref{N}):
\begin{equation}
\kappa =2e^{\phi_c}\sqrt{2\rho_c}.
\end{equation}

These boundary conditions are in accordance with entropy conservation throughout and across the brane: 
\begin{equation}
\label{entropy2}
\dot \rho +(d-1)H(\rho +P)=0.
\end{equation}
Assuming massless thermal radiation up to the critical point, from both sides of the transition, the pressure is given by $P = n^*\Sigma_d T^d$ and the conserved 
thermal entropy by
$S=d (aT)^{d-1}n^*\Sigma_d=$ constant. Then we obtain the following expressions
\begin{equation}
\label{aT}
a_cT_c = aT = {\left(S \over d n^*\Sigma_d\right)}^{1/(d-1)}, ~~{\kappa}=2\sqrt{2(d-1)}~ \sqrt{n^* \Sigma_d}~ ~T_c^{d/2}\,{e^{\phi_c}}
\end{equation}
for the critical value of the scale factor and the relation between the brane tension and the
critical value of the dilaton. 

With massless thermal radiation up to the critical point, we can obtain exact cosmological solutions to the equations of motion.
In the conformal gauge, $N=a$, the string frame scale factor and dilaton are given by \cite{KPT1}
$$
\ln{\left(a \over a_c\right)} ={ 1\over d-2}~\left[ {\eta_+} \ln \left(1+ {\omega a_c|\tau| \over \eta_+} \right) -{\eta_-}  \ln \left(1+{\omega a_c|\tau|\over \eta_-}  \right) \right],
$$
\begin{equation}
\label{solutions}
 \phi=\phi_c+{\sqrt{d-1}\over 2}~\left[ \ln \left(1+ {\omega a_c|\tau|\over \eta_+} \right) - \ln \left(1+ {\omega a_c|\tau|\over \eta_-} \right)  \right], 
\end{equation}
where 
\begin{equation}
4\omega = {\kappa}{d-2\over \sqrt{d-1}},~~~~~ \eta_\pm=\sqrt{d-1}\pm 1.
\end{equation}
The solutions are invariant under time reversal, $\tau \to -\tau$, in accordance with the gluing conditions across the brane
discussed above.
In the neighborhood of the brane, $|\kappa a_c \tau|\ll 1$,
the metric is regular while the dilaton exhibits a conical structure:
$$
\ln{\left(a \over a_c\right)} = {1\over 16(d-1)}\,{(\kappa a_c\tau)^2}+{\cal O}(|\kappa a_c\tau|^3) 
$$
\begin{equation}
 \phi =\phi_c- {{|\kappa a_c \tau|\over 4}}+{\cal O}((\kappa a_c\tau)^2).
\end{equation}
In the far past and future, $|\kappa a_c\tau|\gg 1$, the dilaton asymptotes to a constant,  
the temperature drops and the scale factor tends to infinity. Asymptotically we get 
\begin{equation}
a \sim |\tau|^{2/d-2},
\end{equation}
recovering the characteristic relation between the scale factor and conformal time in a radiation dominated Universe.

The string coupling is bounded by its critical
value at the brane: $g_s \le g_c=e^{\phi_c}$.
Hence our perturbative approach is valid when this critical coupling, $g_c$, is sufficiently small.
The Ricci scalar curvature also attains its maximal value at the brane, which is set by the brane tension: 
\begin{equation}
{\cal R}_c=\kappa^2/4 = {\cal O}(g_c^2).
\end{equation} 
In particular the curvature is finite throughout the cosmological evolution; there is {\it no essential singularity}.
Since the critical coupling sets the value of the brane tension in string units, we conclude that
{\it both $g_s$ and $\alpha'$ corrections remain under control,} provided that $g_c$ is small enough.

We may transform to the Einstein frame via  
\begin{equation}
(N_E, a_E, 1/T_E) = {e^{-{2\phi \over d-2}}}~(N, a, 1/T).  
\end{equation}
Thus the geometrical quantities in the Einstein frame inherit the conical structure of the dilaton field.
The discontinuities of the Hubble parameter in the Einstein frame and the first time derivative of the dilaton (across the transition surface at $T_c$)
 are resolved by the brane pressure, in accordance with the Israel junction conditions \cite{Israel}, which require that the induced metric on the transition surface 
be continuous and the extrinsic curvature to jump by a factor determined by the localized pressure \cite{BKPPT}.

Before we conclude this section, we make a comparison between the string cosmologies we obtained, which incorporate the transition across the stringy
S-brane, with the corresponding solutions in the classical dilaton-gravity theory, coupled to massless thermal radiation, where such a phase
transition is not possible. The latter are given by the solutions in equation (\ref{solutions}) {\it without the absolute value on the time variable,} and coincide
with the stringy solutions in the ``Momentum Regime''. Removing the absolute value on the time variable, 
we see that the dilaton becomes a monotonic function of time, and grows without
a bound in the past. So the
dilaton-gravity system is strongly coupled in the past, and perturbation theory breaks down. 
The string frame scale factor still undergoes a bounce at $\tau=0$, but as we evolve backwards in time,
its first time derivative develops a singularity a finite time in the past, leading to a curvature singularity. 
In the Einstein frame, we recover the singular regime $a_E \to 0$
of classical general relativity. 

The situation is drastically different in the stringy cases. Starting in the contracting phase, the Einstein frame temperature
increases monotonically with time. The analysis of \cite{ADK} reveals that additional thermal states become massless at a critical
temperature determined by the string coupling, ``protecting'' the system from entering the regime $a_E\to 0$ of classical general relativity. 
If at this critical point the coupling is weak, then the additional thermal states are the perturbative winding modes discussed above, and
the analysis of this section applies on how to follow the transition via the stringy S-branes. 
Large enough entropy is needed in order to keep the string frame scale factor parametrically larger
than the string scale, in accordance with equation (\ref{aT}). If on the other hand the coupling at the critical point is large,
then the extra massless thermal states are non-perturbative in nature, but still these can source the required localized pressure needed 
for a reversal to expansion.
Non-perturbative string dualities can be applied in order to map to a weakly coupled description.
It would be interesting to apply such dualities so as to complete the
cosmological history in these cases as well.   

\subsubsection{Spreading the ``Brane Regime''}   
We observe that at the brane, the first time derivative of the temperature vanishes, $\dot T =0$, with the temperature reaching its maximal, critical value $T_c$.  
We explore in this section the possibility of ``spreading'' the ``Brane regime'' 
for an arbitrarily long period of time in the past \cite{KPT2}. During such a long ``Brane regime,'' the string frame temperature remains constant, 
frozen at its critical value.
It follows via the entropy conservation law, equations (\ref{entropy1}) and (\ref{entropy2}), that 
the string frame scale factor must also be constant: $a = a_c$.   
Modulo the running dilaton, such a model shares features with the ``emergent cosmological scenario'' \cite{EM}
where the Universe is in a long quasi-static phase, with $H=0$, before it exits into an expanding phase.

At the critical point, various condensates associated with the extra massless fields may give rise to a dilaton effective potential $V(\phi)$.
In order for the temperature and string frame scale factor to remain constant during the backreacted cosmological evolution, this potential must
take the following form \cite{KPT2}
\begin{equation}
\label{Vd}
V(\phi) = B + Ce^{-2\phi}, ~~ B=P_c;
\end{equation}
that is, the constant part has to be equal to the value of the thermal pressure at the critical temperature, 
while the coefficient of the exponential term can be arbitrary.    
It turns out that the parameters $B$ and $C$ can be obtained in terms of fluxes in the underlying effective gauged supergravity description 
of the thermal system at the extended symmetry point ($\sigma=0$).

Recall that at the critical point, the $U(1)_L$ gauge symmetry associated to the Euclidean time circle gets enhanced to an 
$SU(2)_L$ gauge symmetry. Also, there is at least one spatial circle with radius fixed at the fermionic point, which couples
to the right-moving fermion number $F_R$ and contributes to the spontaneous breaking of the right-moving supersymmetries. 
The $U(1)_R$ gauge symmetry associated to this circle gets enhanced
to an $SU(2)_R$ gauge symmetry.
So at the fermionic extended symmetry point, symmetries reorganize in such a way so that an alternative
description is possible, where
the diagonal $S^1$ cycle blows up to an $SU(2)$-manifold with string-scale volume and flux. 
As a result, at the critical point, the target space can be described in terms of a $d+2$-dimensional space.

Now consider the underlying $d+2$-dimensional gauged supergravity. Once fluxes and gradients along the 2 compact directions are turned on, 
the kinetic terms of the internal massless scalars, graviphotons and matter gauge bosons give rise respectively to the
following dilaton effective potential \cite{KPT2,Roo}:
\begin{equation}
{Ae^{-2\phi}}+{B({\cal U})}+ {{\tilde C}e^{-2\phi}},
\end{equation}  
where ${\cal U}$ stands for the volume of the $10-d$ internal, toroidal manifold.
This form for the effective potential persists when the theory is dimensionally reduced to $d$-dimensions, since the extra 
compact $SU(2)_{k=2}/U(1)$ manifold has fixed volume at the string scale. 
So to support a long ``Brane Regime'' requires to turn on non-trivial fluxes and gradients so that 
\begin{itemize}
\item $B({\cal U})=P_c$
\item $A+ \tilde C = C$, which can be arbitrary.
\end{itemize}  

Thus during a ``Brane Regime'' the only non-trivial evolution is that of the dilaton, which depends on the flux parameter $C$ 
of the effective potential. The only equation that remains to be satisfied is the Friedmann equation, which, for constant scale factor
and dilaton potential given by equation (\ref{Vd}), takes the form: 
\begin{equation}
2\dot\phi^2 = a_c^2\left[(\rho_c +P_c)e^{2\phi}+C\right].
\end{equation}
The solutions corresponding to $C=0$ and $C>0$ are given as follows \cite{KPT2}
$$
{C=0\, : \qquad e^{-\phi}=a_c\,  \sqrt{\rho_c+P_c\over 2}\, (- \tau),\qquad \forall ~\tau \le \tau_+<0,}
$$
\begin{equation}
{C>0\; : \quad e^{-\phi}=\sqrt{\rho_c+P_c\over C}\,\sinh\Big[~a_c\sqrt{C\over 2}(- \tau)\Big],
\quad \forall~ \tau\le \tau_+<0}.
\end{equation}
\begin{itemize}
\item In both cases the ``Brane regime'' starts at  $\tau=-\infty$, with super-weak string coupling.
\item At time $\tau_+ <0$, this regime can exit via a thin S-brane to the ``Momentum, radiation regime,'' described in the 
previous section. The reason for choosing the upper end of the ``Brane Regime'' to be negative is to 
maintain the validity of perturbation theory throughout the cosmological evolution. Notice that throughout the ``Brane Regime'' the
dilaton is monotonically increasing. When it crosses the thin S-brane, it undergoes a bounce and starts to decrease. 
The junction conditions on the first time derivative
of the dilaton just before and after the transition surface can thus be met, as required by a positive tension brane.   
\item The choice of $\tau_+$ determines the critical string coupling at the transition towards the ``Momentum dominated phase'': 
$\phi_+=\phi(\tau=\tau_+)$.
\end{itemize}

In summary, this cosmology can be represented pictorially as follows
\begin{equation}
{{\cal C}(\tau) \equiv \{{\cal B}(\tau\le\tau_c)\} \oplus \{{\cal M} (\tau>\tau_c)\}}.
\end{equation}
Initially the Universe has constant $\sigma$-model 
temperature and scale factor, $T=T_c$ and $a=a_c$.
The string coupling grows from very weak values in the very early past reaching a 
maximal value $g_c$ at $\tau_c$, at which point the Universe exits into the radiation dominated ``Momentum regime''.
Both $g_s$ and $\alpha'$ corrections are under control provided that $g_c$ is small enough.

Finally let us note that the case $C <0$ allows us to obtain a ``Brane regime'' of finite time duration \cite{KPT2}: 
\begin{equation}
{C<0: ~ e^{-\phi}=\sqrt{\rho_c+P_c\over |C|} \,\sin\Big[~a_c\sqrt{|C|\over 2}(-\tau)\Big] \quad 
\tau_-\le \tau\le \tau_+<0},
\end{equation}
opening up the possibility of constructing string cosmological solutions, where an initial ``Winding Regime'' undergoes a transition to
a ``Momentum Regime'' via a thick S-brane. One possibility of realizing this case is by introducing positive spatial curvature \cite{KPT2}, and
so this would be an example of a closed stringy cosmology, which avoids both the Big-Bang and Big-Crunch singularities of classical
general relativity. Of course we would still need to investigate the stability of such a Universe against collapse in the presence of extra matter
and
fluctuations.

\section{Parafermionic cosmologies} 

In this section we consider a class of exact cosmological solutions to classical string
theory, which are
described at the worldsheet level by superconformal field theories (SCFT) of the form \cite{KL,KTT}
\begin{equation}
{SL(2, R)_{-|k|}/U(1)~\times ~K}.
\end{equation}
The first factor corresponds to a gauged Wess-Zumino-Witten model based on the
$SL(2,R)$ group manifold, with the level $k$ taken to be negative. The second factor stands for a suitable internal, compact CFT.
In global coordinates, the sigma-model metric and dilaton field are given by 
$$
 (\alpha')^{-1}{ds^2}= (|k|+2){-dT^2+{dX}^2 \over 1+{T}^2-{X}^2}
$$
\begin{equation}
e^{2\Phi}={e^{2\Phi_0} \over 1+{T}^2-{X}^2 }.
\end{equation}
In the superstring case, this sigma-model metric is exact to all orders in the
$\alpha'$ expansion \cite{Tseytlin}. Additional work on these types of models includes \cite{WZW}.

\begin{figure}
\centering
\includegraphics[width=5cm]{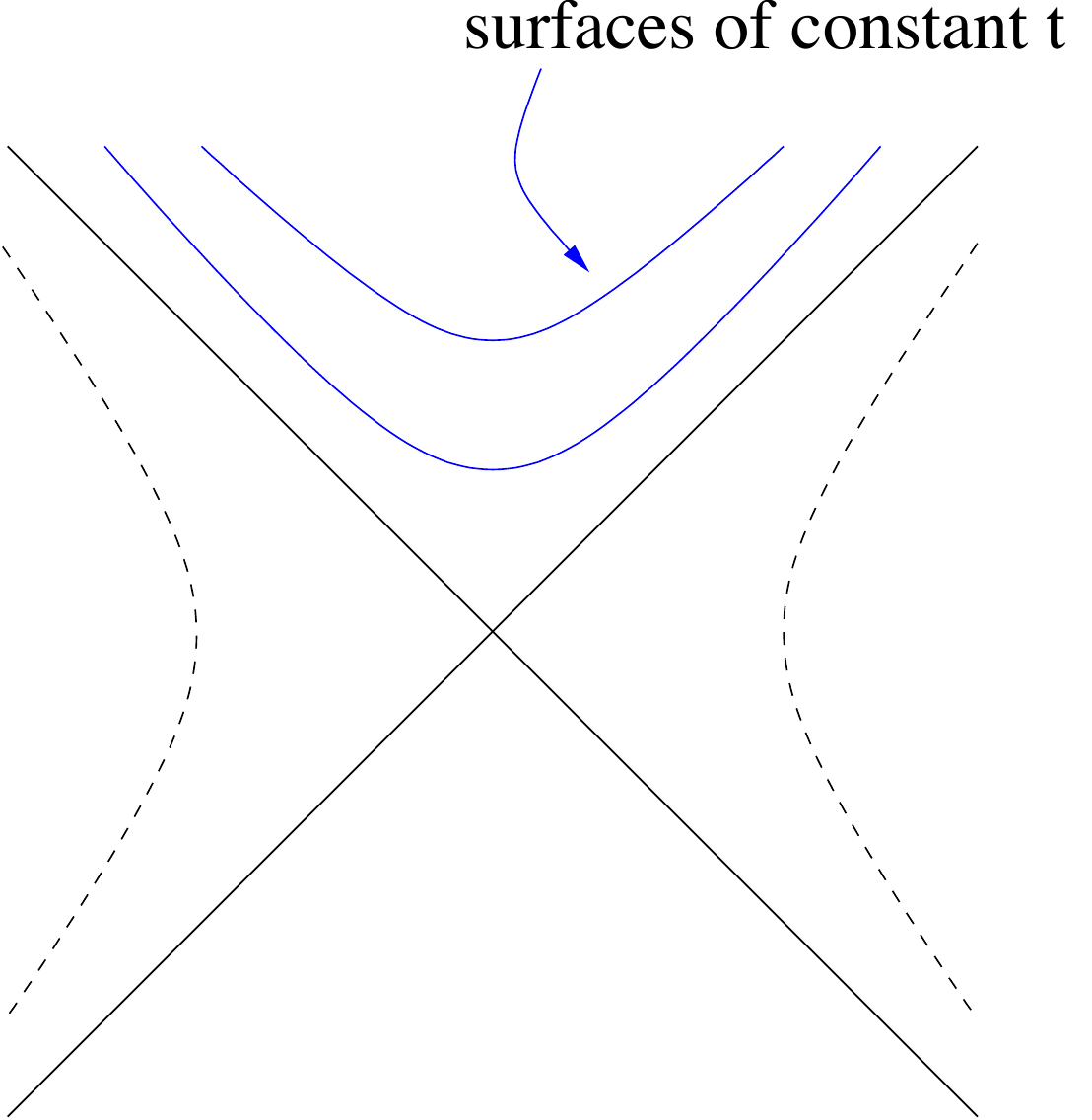}
\caption{The $SL(2,R)/U(1)$ cosmology.}
\label{sl2R}
\end{figure}

The geometry consists of a singularity-free light-cone
region, and there are time-like curvature singularities in the
regions outside the light-cone horizons, where the dilaton field is also singular. See figure \ref{sl2R}. 
The singularities occur at the hyperbolas 
\begin{equation} 
X = \pm \sqrt{1+T^2}.
\end{equation}

\noindent Despite the fact that the singularities follow accelerated trajectories, their proper distance (as measured by the
length of the slices $T=$constant) remains finite
in string units:
\begin{equation}
L=\sqrt{(|k|+2)\alpha'}\int_{-\sqrt{1+T^2}}^{\sqrt{1+T^2}} {dX \over \sqrt{1+T^2-X^2}}=\pi\sqrt{(|k|+2)\alpha'}.
\end{equation}
It can be kept parametrically larger than the string scale for large level $|k|$. 
So in a sense we can think of the cosmology as being spatially closed \cite{KTT}.
Observe also that if we perform a double analytic continuation on the coordinates, 
we obtain Witten's $2$-dimensional black hole solution \cite{WittenBH}.
This double analytic continuation is equivalent to changing the sign of the level $k$ and rotating the Penrose
diagram of the cosmology by $90$ degrees.

At the singularities the sigma-model geometrical description breaks down, but as we will see, 
the underlying worldsheet CFT gives a prescription to describe them.
The cosmological region of interest is the future part of the lightcone region. Parametrizing this region with a 
new set of coordinates $(t,x)$, defined via
\begin{equation}
T=t\cosh x,~~~~X=t\sinh x,
\end{equation}
the metric and dilaton are given by 
\begin{equation} 
(\alpha')^{-1}ds^2=(|k|+2) {-dt^2+ t^2{dx}^2 \over
1+t^2},~~~~~
e^{2\Phi}={e^{2\Phi_0} \over 1+t^2}.
\end{equation}
Therefore we obtain an
expanding, asymptotically flat geometry, with the string coupling
vanishing at late times. Asymptotically we get a timelike linear dilaton background. The surfaces of constant time $t$ are
shown in figure \ref{sl2R}.

The cosmological observer 
never encounters the singularities, as these lie behind
the visible horizons at $T=\pm X$. 
However, signals from the
singularities can propagate into the lightcone region, through the surface $t=0$, and 
influence its future evolution.
In this sense, the cosmology is similar to a Big-Bang cosmology.

The central charge of the
superconformal $SL(2,R)/U(1)$ model at negative level $k$ is given
by  
\begin{equation}
c= 3-{6 \over |k|+2},~~~ \hat{c}=2 -{4 \over |k|+2}. 
\end{equation}
In superstring theory we must tensor it with other conformal field theories so as to satisfy
the condition $\hat c_{\rm tot}=10$ for worldsheet gravitational anomalies to cancel.
To obtain an effectively four dimensional cosmological model, we add two large (however compact) supercoordinates $(y,\, z)$, with 
radii $R_{y, \, z}=R\gg l_s$, together with 
a compact
SCFT system of central charge 
\begin{equation}
\delta \hat{c}= 6+4/(|k|+2).
\end{equation}
Since the cosmological region of interest is non-compact (and asymptotically flat), the model admits the desired four dimensional interpretation \cite{KTT}. 
{\it This is so irrespectively of how small the level $|k|$ is.} As an example consider the case
$|k|=2$. Then we may set the internal CFT factor to be 
\begin{equation}
K \equiv T^2_{y,\, z} \times T^7, 
\end{equation}
where the volume of the $7$-dimensional toroidal manifold is chosen to be at the string scale.
Since the central charge of the $SL(2,R)/U(1)$ factor is smaller than the central charge corresponding to two flat
macroscopic directions, the models are in fact
super-critical.

The metric in the Einstein frame is given by 
\begin{equation}
ds_E^2=(|k|+2)\alpha'( -dt^2+t^2{dx}^2)+ R^2~(1+t^2)(dy^2+dz^2),
\end{equation}
thus describing an anisotropic cosmology. At late times however,
and for large $R$, it asymptotes to an isotropic flat
Freedmann model with 
scale factor growing linearly with time: $a \sim t$.
The effective energy density and pressure supporting this kind of cosmological evolution must satisfy the
following equation of state: 
\begin{equation}
\rho_{\rm eff}=-3P_{\rm eff}.
\end{equation}
Therefore the $SL(2,R)/U(1)$ cosmology is at the cross-point between accelerating and decelerating Universes.

Upon rotation to Euclidean signature,  
we obtain a disk 
parametrized by a radial coordinate $\rho \le 1$ and an angular variable $\phi\in [0,\, 2\pi)$. 
The Euclidean metric and dilaton are given by
\begin{equation}
(\alpha')^{-1}ds^2=(|k|+2)
{d\rho^2+\rho^2d\phi^2 \over 1-\rho^2},~~~~~~ e^{2\Phi}={e^{2\Phi_0}\over 1-\rho^2},
\end{equation}
with the singularity occurring at the
boundary circle $\rho=1$.
The characteristic feature of this geometry is that the radial distance of the center to the boundary of the disk is
finite, but the circumference of the boundary circle at $\rho=1$ is
infinite. Geometrically the space looks like a bell.


The Euclidean background admits a superconformal field theory description  
based on the $SU(2)/U(1)$ coset model at level $|k|$. See e.g. \cite{MMS} for a review.
The interesting feature is that the Euclidean CFT is {\it compact} and unitary.
In fact since it corresponds to an
$N=2$ minimal model, the level $|k|$ must be quantized. So in order for the
Euclidean theory to be well defined, we set $|k|$ to be a positive integer.

This worldsheet CFT is perfectly
well behaved around $\rho=1$, where the geometrical sigma-model description
breaks down.
Near this region the gauged WZW action is given by
\begin{equation}
S_{WZW}=-{|k|+2\over
2\pi}\int d^2z \phi F_{z\bar z}+\dots, 
\end{equation}
with the leading term in the expansion corresponding to a simple topological theory.
From the form of the action near $\rho=1$, we also learn that worldsheet instantons for which, $\int F_{z \bar z}=2\pi i n$, break 
the $U(1)$ symmetry shifting the angle $\phi$ to a discrete symmetry
${Z}_{|k|+2}$, in accordance with the algebraic description of the worldsheet system in terms of $Z_{|k|+2}$ parafermionic currents \cite{FZ}. It
is clear that in the presence of such instantons, the Euclidean path integral is invariant only under discrete shifts of the angle
$\phi$: $\delta \phi =2\pi m/(|k|+2)$.

We argue now that the non-singular description of the theory is an almost-geometrical 
one \cite{KTT}, in terms of a compact T-fold \cite{Tfold}.
In order to construct this, we first perform a T-duality transformation along the angular direction
$\phi$.
The resulting sigma model is based on the following metric and dilaton 
$$
(\alpha')^{-1}ds'^2={(|k|+2)\over 1-\rho'^2}
\left(d\rho'^2+{\rho'^2 \over (|k|+2)^2}d\phi'^2\right) 
$$
\begin{equation}
e^{2\Phi'}={e^{2\Phi_0}\over (|k|+2)(1-\rho'^2)},~~~~\rho'=(1-\rho^2)^{1\over 2}.
\end{equation}
The transformation on the radial coordinate $\rho$ exchanges the boundary of the disk and its center.
The T-dual description is {\it weakly coupled} near $\rho=1$ or $\rho'=0$, where the original
metric and dilaton field were singular.
The only singularity there is a benign orbifold singularity. 
In fact the T-dual description is equivalent to the $Z_{|k|+2}$
orbifold of the original model \cite{MMS,FMS}. Despite the presence of a conical singularity in
the geometry, the string theory amplitudes are finite and well behaved.

The T-fold can now be constructed as follows \cite{KTT}. We take the original disk and cut-off the region past
a non-singular circle, e.g. the region past the circle $\rho =\rho'= 1/\sqrt{2}$, 
$\rho > 1/\sqrt{2}$, containing the apparent singularity at $\rho=1$. 
The cut-off region is replaced with the interior of the
T-dual geometry, $\rho' < 1/\sqrt{2}$, with a well behaved geometrical description (in terms of the
T-dual variables). The two patches are glued together along this non-singular circle via
a T-duality transformation. In particular the Euclidean T-fold is compact, and 
it does not have any boundaries or singularities.
We emphasize that the gluing of the T-dual patches is {\it non-geometrical,} 
as it involves a T-duality transformation on the metric and other fields. 

\begin{figure}
\centering
\includegraphics[width=10cm]{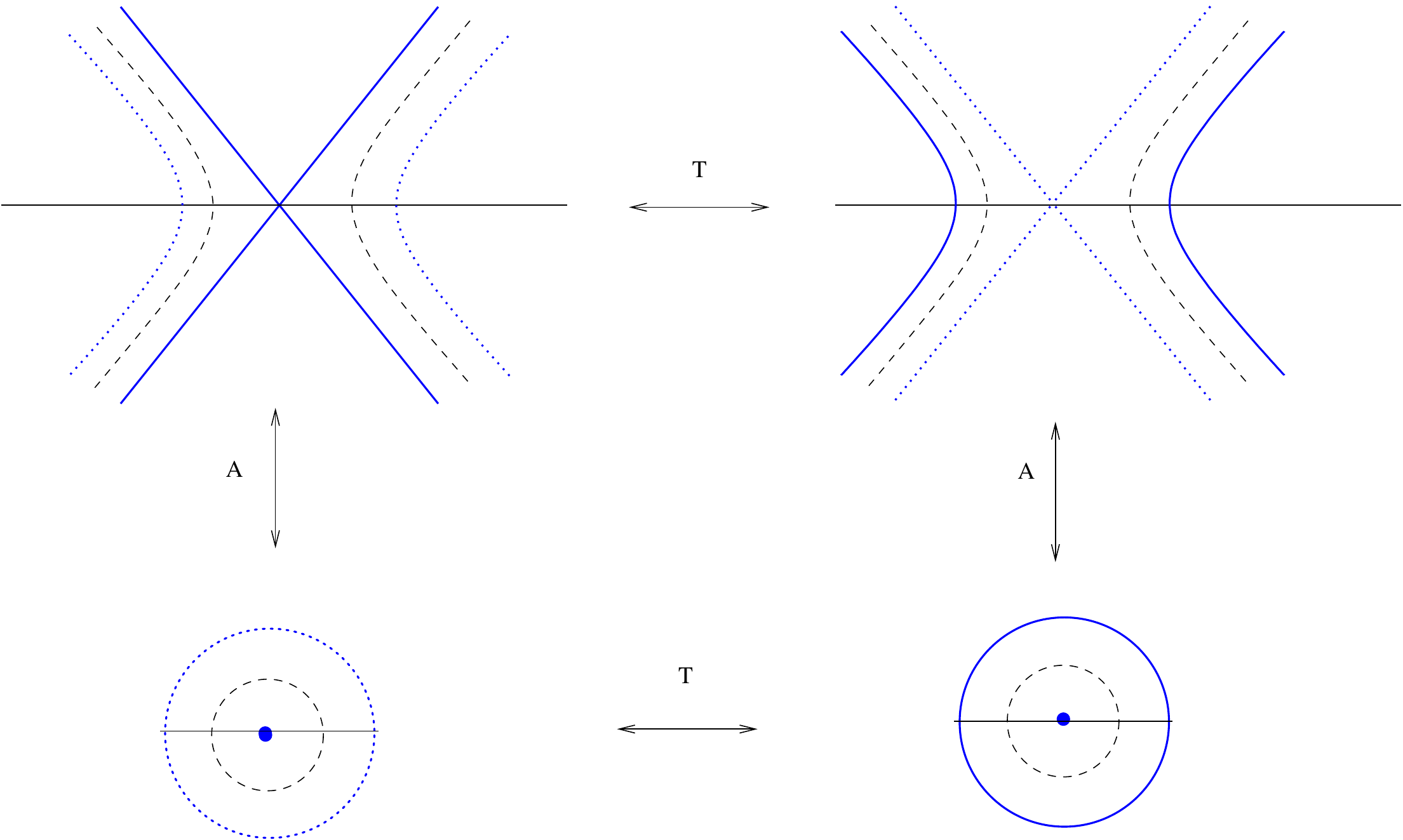}
\caption{The cosmological T-fold. As we cross the hyperbola (striped black line) in the original patch moving towards the apparent
singularity (dotted blue line), we pass to the T-dual patch, and continue motion towards the non-singular lightcone region.}
\label{cosmologyanddisctfold}
\end{figure}

For the cosmology as well, we can obtain
a regular T-fold description as the target space of the CFT.
Here we observe that T-duality interchanges the light-cone and the
singularities \cite{KL}.
We must glue the T-duals along a hyperbola in between the lightcone and the singularities.
The gluings are shown in figure \ref{cosmologyanddisctfold}. The resulting almost-geometrical
description is very similar to $2$-dimensional de-Sitter space, which we can think of as a
hyperboloid embedded in three-dimensional flat space.

We may think of the Euclidean T-fold (or the corresponding compact CFT) as describing a
string field theory instanton, similar in fact to the case of the round sphere which is an
instanton solution to Einstein's classical theory of general relativity in the presence of a positive
cosmological constant. In the latter case, the radius of curvature of the sphere is set by the cosmological constant, while
in the string theoretic case the effective radius of curvature is set by the quantized level $k$. 

\begin{figure}
\centering
\includegraphics[width=6.5cm]{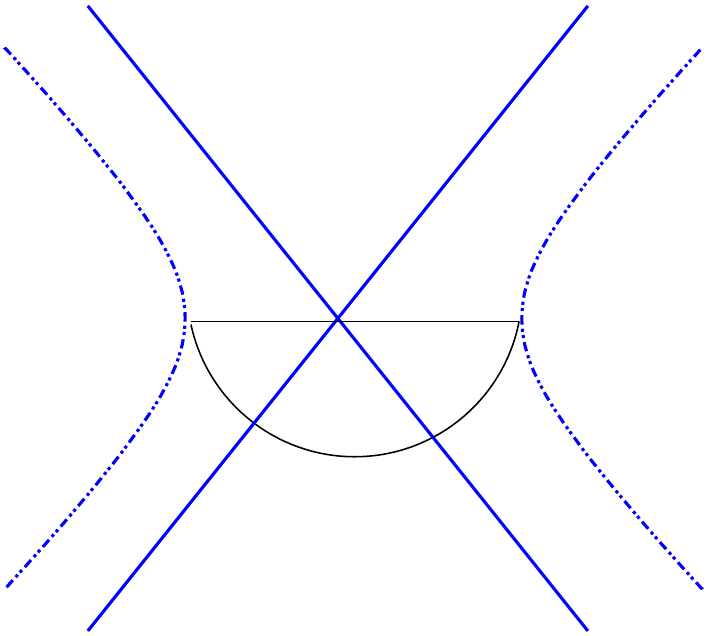}
\caption{The gluing of the half-disk to the cosmology.}
\label{cosmologyanddisc}
\end{figure}

The T-fold construction allows us to {\it formally} define a Hartle-Hawking wavefunction \cite{HH} for the
cosmology, which
we can think of as a vector in the Hilbert space of the 2nd quantized string field theory.
To this extend, we need to perform a ``half T-fold'' Euclidean
path integral over {\it fluctuations of all} target space
fields with specified values on the
boundary \cite{KTT}:
\begin{equation}
\Psi [ h_\partial ,\phi_\partial,\dots ] = \int [dg]
[d\phi] \dots e^{-S(g, \phi, \dots)}.
\end{equation}
More explicitly, we cut the cosmological T-fold in two along a slice of time reversal symmetry, the
$T=0$ slice in each patch, and replace the past geometry with its Euclidean counterpart. Values for the target space fields
must be specified on this $T=0$ slice. See figure \ref{cosmologyanddisc}.  
No other condition needs to be specified since the full Euclidean T-fold has no boundaries
or singularities, thus generalizing the Hartle-Hawking no boundary proposal to string field theory.

The wavefunction is hard to compute, but we can understand some of its global properties
by computing its norm. The norm of the wavefunction is given in terms of the full Euclidean path
integral, over fluctuations around the on-shell closed string background. The full path integral can be computed 
in a 1st quantized formalism by summing over all closed worldsheet 
topologies, including a sum over disconnected diagrams.
In fact it is equal to the exponential of the {\it total, connected} string partition function $Z_{\rm string}$,  
\begin{equation}
||\Psi||^2=e^{Z_{\rm string}}, 
\end{equation}
calculable perturbatively as a sum over all 
(connected) closed Riemann surfaces of genus $0,1,2,\dots,$  
if the string coupling $e^{2\Phi_0}$ is small enough.
Two crucial conditions for the norm to be
finite are that the underlying SCFT be compact and to lead to a {\it tachyon free} Euclidean model \cite{KTT}. In the superstring 
case the latter condition can be satisfied by imposing a suitable GSO projection \cite{G}.

As in the case of de Sitter space, the norm of the wavefunction can be 
interpreted as a thermal ensemble. In the de Sitter case, the norm is given in terms of
the cosmological constant $\Lambda$ by
\begin{equation}
||\Psi_{\rm dS}||^2=e^{3/(8G^2\Lambda)},
\end{equation}
with $3/8G^2\Lambda$ corresponding to the finite de Sitter entropy \cite{HH}.
Likewise in the case of the stringy cosmology at hand, $Z_{\rm string}$ corresponds to a thermal string amplitude \cite{KTT}.
But unlike the de Sitter case, the genus-zero contribution to $Z_{\rm string}$ vanishes.
This is because when the underlying CFT is compact, the spherical CFT partition function is finite and we must
divide it with the infinite volume of the conformal Killing group in order to get the string theory result.
Also since the on-shell background satisfies the sigma-model $\beta$-function equations, its tree-level action
vanishes. As a result in perturbative string theory, the leading contribution arises at the genus-$1$ level.

The effective physical temperature of the $SL(2,R)/U(1)$ cosmological model is set by the
inverse of a radius, which depends on the level $k$ as follows \cite{KTT,TT}:
\begin{equation}
T= 1/2\pi R,~~~~~R=\sqrt{(|k|+2)\alpha'}.
\end{equation}
So it is below the Hagedorn temperature, $R>R_H=\sqrt{2\alpha'}$, for all $|k|> 0$. As a result, the corresponding genus-$1$ thermal string amplitude  
is finite and $\Psi$ is normalizable.
For example, when $|k|=2$, the parafermionic factor of the cosmology has central charge $\hat c=1$, and it is equivalent to a 
fermion together with a compact boson 
at radius $2\sqrt{\alpha'}>R_H$. One way to see that this is the correct value for the radius is that 
for this radius, the T-dual is equivalent to the $Z_4$ orbifold of the original model. 
Explicit computations of the one-loop thermal string partition function in this case can be found in \cite{KTT}. 
Observe that the effective temperature becomes equal to Hagedorn,
$R=R_H$, at $|k|=0$, precisely when the cosmology disappears from the target space.

Hence the norm of the wavefunction $||\Psi||^2$ is finite, and in fact it is a function of the moduli associated to 
the internal CFT $K$ \cite{KTT}.
It would be very interesting to use it in order to define relative probabilities for different string compactifications, related
to these parafermionic cosmologies, as well as precise stringy observables associated to the cosmology. To this end, notice that
in the asymptotically flat region of the cosmology, we can define scattering states, and so an ``S-vector'' can be in turn defined, 
in terms of
the overlaps of these scattering states with the Hartle-Hawking state associated to the cosmology.    

\section{Conclusions}
In this lecture we have uncovered cosmological implications of certain {\it stringy gluing mechanisms}, realizable in a
large class of thermal type II ${\cal N}=(4,0)$ models, which connect distinct 
string effective theories.
The mechanisms are triggered by additional thermal states, which become massless 
at a critical, maximal temperature $T_c$. 
The region around the critical temperature admits a ``brane interpretation,'' with the brane tension 
sourced by non-trivial spatial gradients associated to the extra massless thermal scalars.
In the Einstein frame, the cosmological solutions describe bouncing Universes, connecting in some of the examples
two asymptotically cold thermal phases.
Unlike many versions of pre Big Bang models in the existing literature,
these cosmological solutions remain perturbative throughout the evolution, 
provided that the critical value of the string coupling at the brane is sufficiently small.  
Indeed this class of bouncing cosmologies provides the first examples, where 
{\it both the Hagedorn instabilities as well as the classical Big Bang singularity are successfully resolved,} 
remaining in a perturbative regime throughout the evolution. 

Eternal, bouncing cosmologies open new perspectives, 
to address the cosmological puzzles of standard hot
Big Bang cosmology.
Most of these problems are based on the assumption that the Universe starts out very small 
and hot, with Planckian size and temperature.
In our set up, however, the minimal size of the Universe can be parametrically larger than the string or Planck
scale. The horizon problem in particular is essentially nullified. Causal contact over large scales is assured given the 
fact that the Universe was in a contracting phase for an arbitrarily long period of time.
The large entropy problem does not arise either. If the Universe begins cold and large (larger than the present-day Universe), 
it will by dimensional analysis be likely to contain a sufficient amount of entropy.
The stringy cosmologies we obtained open up the possibility to study the homogeneity problem within a concrete set-up.
Indeed the study of the growth and propagation of cosmological perturbations from the contracting to the expanding phase 
via the stringy S-brane is currently underway \cite{BKPPT}, opening up the possibility of realizing the ``Matter Bounce''
scenario in string theory that produces a scale invariant spectrum of primordial cosmological fluctuations \cite{FB}.
If successful these stringy bouncing cosmologies can form a basis, alternative to inflation, of realizing 
phenomenologically viable models with complete, singularity free cosmological histories.  

\section*{Acknowledgement}
We are grateful to the organizers of the {\it Corfu Summer Institute 2012} for inviting us to present this lecture. 
This work is based on research done in collaboration with C. Angelantonj, R. Brandenberger, I. Florakis, H. Partouche, S. Patil and J. Troost. 
The work of C.K. 
is partially supported by the ANR 05-BLAN-NT09-573739 and a PEPS contract. The work of C.K. and N.T. is supported 
by the CEFIPRA/IFCPAR 4104-2 project and a PICS France/Cyprus.

\end{document}